\documentclass[preprintnumbers,showpacs,amsmath,amssymb,floatfix,9pt,prd,onecolumn,
superscriptaddress,nofootinbib]{revtex4}
\usepackage{latexsym}
\usepackage{epsfig}
\usepackage{amssymb}

\begin{document}

\title{Dark Energy Stars in Tolman-Kuchowicz spacetime in the context of Einstein Gravity }

\author{Piyali Bhar}
\email{piyalibhar90@gmail.com, piyalibhar@associates.iucaa.in}
\affiliation{Department of Mathematics, Government General Degree College, Singur,\\
Hooghly, West Bengal 712 409, India}

\begin{abstract}

Dark energy is the component in the present Universe with the greatest abundance, and it is responsible for the accelerating expansion of the Universe. As a result, dark energy is likely to interact with any compact astrophysical object [Muhammad F.A.R. Sakti and Anto Sulaksono, {\it Phys. Rev. D} {\bf 103}, 084042 (2021)]. In present paper, we propose a model for a dark energy star made up of dark and ordinary matter in which the density of dark energy is proportional to the density of isotropic perfect fluid matter. In the context of general relativity, the model is derived in the curved Tolman-Kuchowicz spacetime geometry [Tolman, Phys Rev 55:364, (1939); Kuchowicz, Acta Phys Pol 33:541, (1968)]. Here, we look at how dark energy affects stellar mass, compactness, and equilibrium etc. The physical parameters of the model e.g., pressure, density, mass function, surface redshift etc. are investigated, and the stability of stellar configuration is studied in detail. The model has interesting properties because it meets all energy criteria and is free from central singularities. The maximum allowable mass has been obtained from our model with the help of $M-R$ diagram. We analyse  many physical properties of the model and checked that it meets all regularity constraints, is stable, and therefore physically realistic.

\end{abstract}

\maketitle

\section{Introduction}
In cosmology, observed matter in the Universe reports with only 5\% of observed gravity, with the remaining 26\% and 69\% accounted for it through dark matter and dark energy, respectively \cite{Planck:2015fie}. Dark energy is evenly dispersed throughout the universe, not only in space but also in time, so its impact is not diluted as the universe expands. The actual nature of both of the dark phenomena is unknown, and the quest for the missing mass of the Universe has even led to significant changes to Newton's and Einstein's gravity theories \cite{Ferreira:2009eg}. Unlike both kinds of matter, dark energy is generally uniform in time and space and gravitationally repellent within the volume it occupies. The origins of dark energy are still a mystery. Dark energy can be found by looking at how fast the cosmos expands and how quickly large-scale structures like galaxies and clusters of galaxies emerge due to gravitational instabilities. The use of telescopes to measure the distance of objects seen at different size scales in the history of the universe is required to determine the expansion rate. Nevertheless, every experimental test has completely validated our understanding of general relativity \cite{LIGOScientific:2016emj}. Precision measurements of the cosmic microwave background (CMB) have revealed that the overall energy density of the universe is extremely close to the critical density required to make it flat.
Recent astrophysical observations have proven that the expansion of the Universe is accelerating. Measurements of type Ia supernovae \cite{Riess:2001gk,Perlmutter:1999jt}, the galaxy rotation curves and microwave background radiation \cite{Bennett:2003bz,Hinshaw:2003ex} have revealed evidence of this expansion. The existence of so-called dark energy is one of the assumptions of the present cosmological paradigm, where general relativity is assumed to be valid. The dark energy component is characterized by a negative pressure. Since dark energy opposes gravity, it accelerates the expansion of the universe while delaying the formation of large-scale structure. This negativity corresponds to the violation of the strong energy condition (SEC). The cosmological constant, which represents a constant energy density filling space uniformly, and scalar fields, such as quintessence or moduli, which are dynamic quantities with energy densities that can fluctuate in time and space, are two hypothesized forms of dark energy. The cosmological constant usually includes contributions from scalar fields that are constant in space. If dark energy is linked to Einstein's constant $\Lambda$, the local density of dark energy must be the same as that found in global observations \cite{SupernovaSearchTeam:1998fmf,SupernovaCosmologyProject:1998vns}. Since dark energy fills up most of our universe, it may interact any ordinary matter within any local astrophysical phenomenon in wormholes, black holes, compact stars, and other strange compact objects \cite{Sakti:2021mvd}. Because of the basic importance of dark energy in cosmology, we naturally look for local astrophysical manifestations of it. The equation of state (EoS) of dark energy can be described by $p = \omega \rho$ with $\omega < -1/3$. Kiselev \cite{Kiselev:2002dx} constructs the first black hole solution with such an equation of state (EoS), which is termed a quintessential black hole. Refs. \cite{Fernando:2012ue,Xu:2016jod,Ghosh:2015ovj,Sakti:2019udk,Sakti:2019krw} contain extended solutions of this classic black hole, as well as studies of them that include quantum characteristics. Furthermore, in \cite{Bhar:2016vdn}, wormholes sustained by dark energy are investigated by using $\omega <-1$ to represent phantom dark energy. Although the EoS can be expressed in the manner $p = \omega \rho$, such compact objects may have various EoSs but negative principal pressures if SEC is still violated. Chernin et al. \cite{1271C} combine the modified Kahn-Woltjer method for the Milky Way-M 31 binary and HST observations of the expansion flow around the Local Group to study the local density of dark energy and the dark matter mass contained within the Local Group analysis in a self-consistent way and simultaneously provided new strong evidence in favour of Einstein's idea of the universal antigravity described by the cosmological constant.\par

In astronomy, the structure of relativistic stars and the phenomenon of gravitational collapse are of immense interest of research. Tolman \cite{Tolman:1939jz} developed a method for providing explicit solutions for static spheres of fluid, which proved useful in the study of stellar structure. Schwarzschild \cite{Schwarzschild1916} considered analytic solutions describing a star of uniform density. Using the equation of state for a cold Fermi gas, Oppenheimer and Volkoff \cite{Oppenheimer:1939ne} explored the gravitational equilibrium of neutron stars by suitably choosing specific Tolman solutions. In this connection, one may consider Wyman's work \cite{Wyman1}, in which isotropic coordinates were utilised to solve the relativistic equations of a perfect fluid with constant energy density. A critical analysis and generalisation of the Tolman solutions were described in \cite{Wyman2}. There have been a number of investigations on compact astrophysical objects in which the energy density $\rho$ and pressure $p$ obey an equation of state typical of dark energy, such as $p=-\rho$. In the literature, such items have been given numerous names. For the sake of simplicity, we call them ``dark energy stars". Lobo's concept of a stable dark energy star is based on the assumption of two spatial types of mass functions: one with constant energy density and the other with a Tolman \& Whitker mass function \cite{Lobo:2005uf,Lobo:2006ue}. The system is determined to be stable under a minor linear perturbation, and all of the properties of the dark energy star have been discussed in these papers. Beltracchi and Gondolo \cite{Beltracchi:2018ait} proposed a model of dark energy stars of finite size with a dark energy like internal equation of state. They described the collapse of a spherical system from an initial state of positive pressure to a final state with a dark energy core using a time-dependent solution of Einstein's field equations. There are no singularities, event horizons and no violation of the weak or null energy conditions in the suggested model. Yazadjiev \cite{Yazadjiev:2011sm} constructed a general class of exact interior solutions describing mixed relativistic stars containing both ordinary matter and dark energy in different proportions. According to the author, the phantom field represents dark energy within the star. Sakti and Sulaksono \cite{Sakti:2021mvd} proposed a class of interior solutions of the Einstein field equation with perfect fluid matter and phantom scalar field. Dayanandan and Smitha \cite{Dayanandan:2021tku} proposed  anisotropic DE star by adopting Tolman IV - type gravitational potential and show that all energy conditions are satisfied except strong energy condition. Mustafa et al. \cite{Mustafa:2021ldz} explored the Rastall theory of modified gravity for three different compact stars on the basis of observational facts, by adopting an explicit metric potentials, which meet the Karmarkar condition by adopting the anisotropic fluid. Bhar and Rahaman \cite{Bhar:2014dqa} suggest a new dark energy star model that consists of five zones: a solid core with constant energy density, a thin shell between the core and the interior, an inhomogeneous inner area with anisotropic pressures, a thin shell, and an external vacuum region. In this study, the stability criterion under a small radial perturbation is also explored. Bhar et al. \cite{Bhar:2016dhd}  developed a spherically symmetric stellar configuration based on the assumption that the stellar configuration's matter distribution is anisotropic in nature, by using the dark energy equation of state and compare the result to realistic phenomena, such as low mass X-ray binaries and X-ray pulsars. With three unified equations of state constructed by the Brussels-Montreal group, Smerechynskyi et al. \cite{Smerechynskyi:2020cfu} investigated the density distribution of minimally-coupled scalar field dark energy inside a neutron star. They study how the existence of dark energy affects the macroscopic properties and the value of the mass limit for neutron stars using the predicted density distribution of dark energy inside the star.  By considering negative anisotropic pressures, Rashida Bibi et al. \cite{Bibi:2016rrt} were able to obtain a new class of solutions for the Einstein-Maxwell field equations for static spherically symmetric space-times, which is a potential model of a dark energy star. Das and Ali \cite{Das:2014swa} obtained a model of dark energy star in Krori-Barua space time with Einstein-Maxwell field equations in presence of charged fluids with anisotropic pressures. Halpern and Pecorino \cite{Halpern:2013pga} examined the question of energy localization for an exact solution of Einstein's equations with a scalar field corresponding to the phantom energy interpretation of dark energy. \par
In this paper, we study the models of relativistic stars that contain both ordinary matter and dark energy. Because the existence of dark energy, we can assume that today's stars are made up of a mix of ordinary matter and dark energy in different proportions. The study of such mixed objects is a novel and exciting subject, and some progress has already been made in this direction (see for example \cite{Lobo:2006ue,bronnikov2006regular,Dzhunushaliev:2008bq,Chan:2008rk,Ghezzi:2009ct,Ciarcelluti:2010ji,Rahaman:2011hd,Dzhunushaliev:2011xx,Caldwell:1999ew} and references therein). We have organized our paper as follows: in Sec.~\ref{interior} we have describe the basic field equations in presence of dark energy. The descriptions of the metric coefficients are described in details. The next section describes about the solutions of the field equations and a smooth matching between interior and exterior spacetime. The physical analysis of our present model has been discussed in Sec.~\ref{pa}. The stability and equilibrium conditions of the current model are described in Sec.~\ref{equ} through various tests, and some conclusions are presented in the final section.

\section{Interior spacetime and Field Equations}\label{interior}
The Einstein's field equations are given by,
\begin{eqnarray}
R_{\mu\nu}-\frac{1}{2}g_{\mu \nu}R &=&\frac{8\pi G}{c^4}T_{\mu \nu},
\end{eqnarray}
where $R_{\mu \nu}$ is the Ricci tensor, $R$ denotes the scalar curvature, and $T_{\mu \nu}$ is the energy momentum tensor.\\
The line element in standard Schwarzschild coordinates $x^{\mu}=(t,\,r,\,\theta,\,\phi)$ in $4$D spacetime is described by,
\begin{equation}
ds^{2}=e^{\nu}dt^{2}-e^{\lambda}dr^{2}-r^{2}(d\theta^{2}+\sin^{2}\theta d\phi^{2}),
\end{equation}
In the above equation, the metric potentials $e^{\lambda}$ and $e^{\nu}$ are taken as a function of `r', assuming a static space-time.\\
Let us assume that the energy-momentum tensor is composed of matter with mass-energy density $\rho$ and pressure $p$, as well as dark energy with density $\rho^{de}$, radial pressure $p_r^{de}$, and tangential pressure $p_t^{de}$. In terms of the (variable) cosmological constant the dark energy density is given by $\rho^{\text{de}}=\frac{\Lambda}{8\pi}$ \cite{Ghezzi:2009ct}.\\
We write the corresponding energy-momentum tensor of the two fluids as \cite{Ghezzi:2009ct},
  \begin{eqnarray}
  T_0^0=\rho^{\text{eff}}=\rho+\rho^{de} ,\label{t1}\\
  T_1^1=-p_r^{\text{eff}}=-(p+p_r^{de}),\\
  \text{and}~~~
  T_2^2=T_3^3=-p_t^{\text{eff}}=-(p+p_t^{de}), \label{t3}\\
  T_0^1=T_1^0=0.
\end{eqnarray}
The Einstein field equations are as follows (assuming $G=1=c$):
\begin{eqnarray}
T_0^0:~8\pi(\rho+\rho^{de})&=&e^{-\lambda}\left[\frac{\lambda'}{r}-\frac{1}{r^{2}} \right]+\frac{1}{r^{2}},   \label{fe1}\\
T_1^1:~8\pi (p+p_r^{de}) &=&e^{-\lambda}\left[\frac{1}{r^{2}}+\frac{\nu'}{r} \right]-\frac{1}{r^{2}}, \label{fe2}\\
T_2^2=T_3^3:~8\pi (p+p_t^{de}) &=& \frac{1}{2}e^{-\lambda}\left[ \frac{1}{2}\nu'^{2}+\nu''-\frac{1}{2}\lambda'\nu'+\frac{1}{r}(\nu'-\lambda')\right].\label{fe3}
\end{eqnarray}
where $()^{\prime}$ denotes derivative with respect to the radial co-ordinate `r'.\par
For our present paper we have taken the well known metric potentials proposed by Tolman-Kuchowicz \cite{Tolman:1939jz,Kuchowicz1968} (henceforth TK) as,
\begin{eqnarray}\label{tk1}
\nu(r)= Br^2+2\ln D,\,\lambda(r)= \ln(1 + ar^2 + br^4),
\end{eqnarray}
where $a$, $B$ and $b$ are constants with units km$^{-2}$, km$^{-2}$ and km$^{-4}$, respectively. $D$ a dimensionless constant. We shall calculate the numerical values of these constants from a smooth matching of interior and exterior spacetimes. The metric potentials provide a non-singular stellar model which will be described in the coming sections.\par
Several researchers from the field of general relativity as well as in the field of modified gravity have used the above metric potentials earlier. In connection with the MIT Bag model equation of state, Biswas et al. \cite{Biswas:2019doe} used this metric potentials to obtain a strange star structure. Under Einstein's general theory of relativity, in the presence of the cosmological constant $\Lambda$ (where $\Lambda$=$\Lambda(r)$), Jasim et al. \cite{Jasim2018} examined a unique model for spherically symmetry of anisotropic strange stars. On the other hand, in the context of modified gravity, we now discuss the uses of TK metric potentials. Bhar et al. \cite{Bhar2019a} had previously employed this metric potential to model compact objects in Einstein Gauss Bonnet gravity and Javed et al. \cite{j1} used this metric potential to model anisotropic spheres in $f(R, G)$ modified gravity, Majid and Sharif \cite{majid} obtained
quark stars in massive Brans-Dicke gravity, Biswas et al. \cite{sbiswas2020a} obtained an anisotropic strange star with $f(R,\,T)$ gravity, Farasat Shamir
and Fayyaz \cite{fs} obtained the model of charged compact star in $f(R)$ gravity, Naz and Shamir \cite{naz} found the stellar model in $f(G)$ gravity, Rej et al. \cite{prej2021} studied the charged compact star in the context of $f(R,\,T)$ gravity. One notices that TK metric potential succeed in producing the model of a compact star that lacks singularity and meets all the physically acceptable requirements.\par
With the help of the expressions given in Eqn. (\ref{tk1}), the field equations (\ref{fe1})-(\ref{fe3}) take the following form :
\begin{eqnarray}
8\pi(\rho+\rho^{de})&=&\frac{3 a + (a^2 + 5 b) r^2 + 2 a b r^4 +
 b^2 r^6}{8 \pi (1 + a r^2 + b r^4)^2},   \label{f1}\\
8\pi (p+p_r^{de}) &=& \frac{2B-a -b r^2}{8\pi(1 + a r^2 + b r^4)}, \label{f2}\\
8\pi (p+p_t^{de}) &=&\frac{-a + 2 B + (-2 b + B (a + B)) r^2 + a B^2 r^4 +
 b B^2 r^6}{8\pi(1 + a r^2 + b r^4)^2} .\label{f3}
\end{eqnarray}

To solve the equations (\ref{f1})-(\ref{f3}), by following the previous work done by the authors mentioned in the Refs \cite{Ghezzi:2005iy,Ghezzi:2009ct,Rahaman:2011hd}, let us assume that the radial pressure corresponding to dark energy ($p_r^{de}$)
is proportional to the density corresponding dark energy, i.e.,
\begin{eqnarray}\label{e1}
p_r^{de}=-\rho^{de},
\end{eqnarray}
along with the density corresponding to dark energy is proportional to the normal baryonic matter density, i.e.,
\begin{eqnarray}\label{e2}
\rho^{de}=\alpha \rho,
\end{eqnarray}
where $\alpha$ is a non-zero constant and its value will be obtained from the boundary conditions in the coming section. It is worth noting that this type of equation of state (equations (\ref{e1}) \& (\ref{e2})) means that the matter distribution under consideration is in tension, and hence the matter is referred to as a `false vacuum', or `degenerate vacuum' or `$\rho$ -vacuum'.

\section{Our proposed model of dark energy star}
Solving the field equations (\ref{f1})-(\ref{f3}) with the help of (\ref{e1}) and (\ref{e2}), the matter density and the pressure for the normal baryonic matter are obtained as, \begin{eqnarray}
  \rho &=& \frac{3 a + (a^2 + 5 b) r^2 + 2 a b r^4 +
 b^2 r^6}{8 (1 + \alpha) \pi (1 + a r^2 + b r^4)^2},\\
  p &=& \frac{2 a \alpha +
 4 \alpha b r^2 - (a - 2 (1 + \alpha) B + b r^2) (1 + a r^2 +
    b r^4)}{8 (1 + \alpha) \pi (1 + a r^2 + b r^4)^2}.
\end{eqnarray}
The matter density, radial and transverse pressure due to the dark energy
are obtained as,
\begin{eqnarray}
\rho^{de}&=&\frac{\alpha (3 a + (a^2 + 5 b) r^2 + 2 a b r^4 + b^2 r^6)}{8 (1 +
   \alpha) \pi (1 + a r^2 + b r^4)^2},\\
  p_r^{de} &=& -\frac{\alpha (3 a + (a^2 + 5 b) r^2 + 2 a b r^4 + b^2 r^6)}{
 8 (1 + \alpha) \pi (1 + a r^2 + b r^4)^2},\\
 p_t^{de}&=&\frac{1}{8 (1 + \alpha) \pi (1 + a r^2 +
   b r^4)^2}\Big[-3 a \alpha + (a^2 - (1 + 6 \alpha) b -
    a (1 + \alpha) B + (1 + \alpha) B^2) r^2 \nonumber\\&&+\big(2 a b -
    2 (1 + \alpha) b B + a (1 + \alpha) B^2\big) r^4 +
 b (b + (1 + \alpha) B^2) r^6\Big].
\end{eqnarray}

Now the effective density for our present model is obtained as,
\begin{eqnarray}
  \rho^{\text{eff}} =\rho+\rho^{de}&=& \frac{3 a + (a^2 + 5 b) r^2 + 2 a b r^4 +
 b^2 r^6}{8 \pi (1 + a r^2 + b r^4)^2},
\end{eqnarray}
and the effective radial and transverse pressures are obtained as,
\begin{eqnarray}
  p_r^{\text{eff}}=p+p_r^{de}&=& \frac{2B-a -b r^2}{8\pi(1 + a r^2 + b r^4)},\\
  p_t^{\text{eff}}=p+p_t^{de}&=&\frac{-a + 2 B + (-2 b + B (a + B)) r^2 + a B^2 r^4 +
 b B^2 r^6}{8\pi(1 + a r^2 + b r^4)^2}.
\end{eqnarray}
The values of $a,\,b,\,B$ and $D$ must be fixed in order to construct the profiles of the model parameters. We match our interior spacetime smoothly to the exterior Schwarzschild line element \cite{Schwarzschild1916} at the boundary $r=\mathcal{R}$ to obtain the values of four unknown constants. The following is the exterior spacetime:
\begin{eqnarray}
ds^{2}&=&F(r) dt^{2}-F(r)^{-1}dr^{2}-r^{2}(d\theta^{2}+\sin^{2}\theta d\phi^{2}),
\end{eqnarray}
 where $F(r)=\left(1-\frac{2\mathcal{M}}{r}\right)$ and $\mathcal{M}$ being the mass of the compact star.\\
Now at the boundary $r = \mathcal{R}$ the metric coefficients
$g_{rr}$, $g_{tt}$ , $\frac{\partial}{\partial r}(g_{tt})$ all are continuous, which gives the following set of equations :
\begin{eqnarray}
\text{Continuity of $g_{tt}$ :~}1-\frac{2\mathcal{M}}{\mathcal{R}}=e^{B\mathcal{R}^2}D^2,\label{b1}\\
\text{Continuity of $g_{rr}$ :~}\left(1-\frac{2\mathcal{M}}{\mathcal{R}}\right)^{-1}=1+a\mathcal{R}^{2}+b \mathcal{R}^{4},\\
\text{Continuity of $\frac{\partial}{\partial r}(g_{tt})$ :~}\frac{2\mathcal{M}}{\mathcal{R}^2}=2B\mathcal{R}e^{B\mathcal{R}^2}D^2. \label{b3}
\end{eqnarray}
Solving the Eqns. (\ref{b1})-(\ref{b3}), one can obtain,
\begin{eqnarray}
a&=&\frac{1}{\mathcal{R}^2}\left[\left(1-\frac{2\mathcal{M}}{\mathcal{R}}\right)^{-1}-1-b \mathcal{R}^4\right],\\
B&=&\frac{\mathcal{M}}{\mathcal{R}^3}\left(1-\frac{2\mathcal{M}}{\mathcal{R}}\right)^{-1},\\
D&=&e^{-\frac{B\mathcal{R}^2}{2}}\sqrt{1-\frac{2\mathcal{M}}{\mathcal{R}}}.
\end{eqnarray}
Now using the condition $p(r=\mathcal{R})=0$, we get the value of $\alpha$ as,
\begin{eqnarray}\label{beta}
\alpha&=&\frac{(a - 2 B + b \mathcal{R}^2) (1 + a \mathcal{R}^2 + b \mathcal{R}^4)}{2 (a + B + (2 b + a B) \mathcal{R}^2 +
   b B \mathcal{R}^4)}
\end{eqnarray}
As a result, we have successfully obtained all of the constants presented in the metric potentials in terms of mass and radius of the compact star model.
We calculated the numerical values of various constants in table \ref{table1} using observed values of several candidates for compact stars.

\begin{table*}[t]
\centering
\caption{The numerical values of $a,\,B$ and $D$ for some well known compact objects by assuming $b=0.5\times 10^{-5}$~km$^{-4}$.}\label{table1}
\begin{tabular}{@{}ccccccccccccc@{}}
\hline
Star & Observed mass & Observed radius & Estimated  & Estimated &  $a$&$B$&$D$\\
& $M_{\odot}$ & km. & mass ($M_{\odot}$) & radius (km.)& $km^{-2}$ & $km^{-2}$\\
\hline
Her X-1 \cite{Abubekerov:2012yj}& $0.85 \pm 0.15$ & $8.1 \pm 0.41$ & 0.85 & 8.5 & 0.00543030 & 0.00289578& 0.756246       \\
SMC X-4 \cite{Rawls:2011jw} & $1.29 \pm 0.05$ & $8.831 \pm 0.09$ & 1.29 & 8.8 & 0.00945188 & 0.00491954 & 0.622699 \\
Vela X-1 \cite{Rawls:2011jw} & $1.77 \pm 0.08$ & $9.56 \pm 0.08$    & 1.77 & 9.5& 0.01307120 & 0.00676124 & 0.494630     \\
4U 1538-52 \cite{Rawls:2011jw}& $0.87 \pm 0.07$ & $7.866 \pm 0.21$ & 0.87 & 7.8 & 0.00775626 &0.00403023&0.724610  \\
LMC X-4 \cite{Rawls:2011jw} & $1.04 \pm 0.09$ & $8.301 \pm 0.2$ & 1.04 & 8.3 & 0.00816755& 0.00425600 & 0.685691  \\
Cen X-3 \cite{Rawls:2011jw} & $1.49 \pm 0.08$ & $9.178 \pm 0.13$ & 1.49 & 9.2& 0.01038580& 0.00540449& 0.574909 \\
PSR J1614-2230 \cite{Demorest:2010bx} & $1.97 \pm 0.04$ & $9.69 \pm 0.2$   & 1.97& 9.7& 0.01541370 & 0.00794205 &0.435751 \\
PSR J1903+327 \cite{Freire:2010tf} & $1.667 \pm 0.021$ & $9.438 \pm 0.03$ & 1.67 & 9.4 &0.01202160&0.00623168&0.523832           \\
4U 1820-30 \cite{Guver:2008gc} & $1.58 \pm 0.06$&  $9.316 \pm 0.086$  & 1.58 & 9.3&0.01118440&0.00580843&0.549392  \\
EXO 1785-248 \cite{Ozel:2008kb} & $1.3 \pm 0.2$ & $8.849 \pm 0.4 $ &1.4 &8.85&0.01078010&0.00558588&0.586810 \\
\hline
\end{tabular}
\end{table*}

\begin{table*}[t]
\centering
\caption{The numerical values of $\alpha$, central density ($\rho_c$), surface density ($\rho_s$), central pressure ($p_c$), compactness ratio ($\mathcal{U}$) and surface redshift ($z_s$), $p_c/\rho_c$ have been shown for different compact stars by assuming $b=0.5\times 10^{-5}$~km$^{-4}$.}\label{table2}
\begin{tabular}{@{}ccccccccccccc@{}}
\hline
Star & $\alpha$&$\rho_c$&$\rho_s$&$p_c$&$\mathcal{U}$& $z_s(r_b)$& $p_c/\rho_c$\\
& & gm.cm$^{-3}$ & gm.cm$^{-3}$ & dyne.cm$^{-2}$ \\
\hline
Her X-1& $-2.97284\times10^{-17}$& $8.74626 \times 10^{14}$& $5.47576\times 10^{14}$ & $1.74553\times 10^{34}$ & 0.147500 & 0.19098 &0.0221749\\
SMC X-4 & $-3.00195\times10^{-17}$& $ 1.52236 \times 10^{15}$ &$6.53513\times 10^{14}$ & $1.87092\times 10^{34}$ & 0.216222 & 0.32738 & 0.0136551\\
Vela X-1 & $-7.50627\times10^{-17}$&$ 2.10530\times 10^{15}$&  $6.31302\times 10^{14}$ & $2.18040\times 10^{34}$ &0.274816&0.49010&0.0115075\\
4U 1538-52 &$3.88055\times10^{-17}$& $1.24925\times 10^{15}$&  $6.94702\times 10^{14}$ & $1.46987\times 10^{34}$ & 0.164519& 0.22082&0.0130733\\
LMC X-4& $-2.4061\times10^{-16}$&$1.31559\times 10^{15}$&  $6.65943\times 10^{14}$& $1.66435\times 10^{34}$ & 0.184819 & 0.25952&0.0140577\\
Cen X-3& $-9.21681\times 10^{-17}$&$1.67277\times 10^{15}$&  $6.31976\times 10^{14}$& $2.04487\times 10^{34}$ & 0.238886 &0.38379&0.0135827  \\
PSR J1614-2230 &$-1.45832\times10^{-16}$&$2.48258\times 10^{15}$&  $6.24067\times 10^{14}$& $2.27317\times 10^{34}$ & 0.299562 & 0.57941&0.0101739 \\
PSR J1903+327 &$3.76492\times10^{-17}$&$1.93624\times 10^{15}$&  $6.32284\times 10^{14}$& $2.13474\times 10^{34}$ & 0.262048 & 0.44957 &0.0122502 \\
4U 1820-30   &$4.00604\times10^{-17}$&$1.80140\times 10^{15}$&  $6.33028\times 10^{14}$& $2.08956\times 10^{34}$ & 0.250591 & 0.41589&0.0128885 \\
EXO 1785-248 &$3.18101\times10^{-17}$&$1.73629\times 10^{15}$&  $6.73061\times 10^{14}$& $1.89224\times 10^{34}$ & 0.233333& 0.36931&0.0121091\\
\hline
\end{tabular}
\end{table*}

\section{Physical analysis}\label{pa}
We shall look at several physical characteristics of relativistic compact star formations in this section. We have considered the compact star 4U 1538-52 as an example, which has a mass of $0.87 M_{\odot}$ and a radius of $7.8$ km. Along with this information, we will plot various model parameters and investigate various physical aspects in order to obtain more accurate stellar configurations.

\subsection{Regularity of physical parameters}
\begin{itemize}
\item {\bf Nature of metric potential} :~Within the radius of the star, both the metric potentials have no singularities. Moreover for our present stellar model, $e^{\nu(0)}=D^2$, a non-zero constant, and $e^{-\lambda(0)}=1$. The derivative of the metric coefficients give $(e^{\lambda})'=2ar+4br^3$, $(e^{\nu})'=2BD^2re^{Br^2}$. The derivative of the metric potentials are $0$ at the centre of the star. They are also positive and consistent within the interior of the star. Fig.~\ref{metric} depicts the profile of the metric coefficients. At the boundary, the interior metric potentials are smoothly matched to the metric components of the exterior Schwarzschild line element.\par

\begin{figure}[htbp]
    \centering
        \includegraphics[scale=.55]{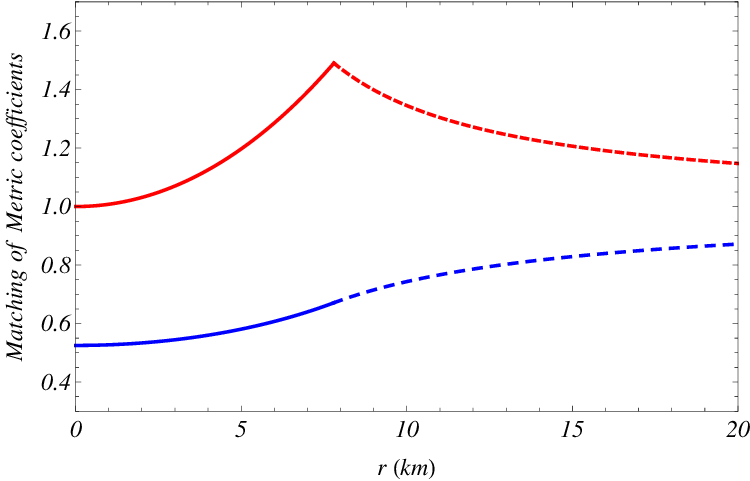}
       \caption{The matching condition of the metric potential $e^{\lambda}$ (red) and $e^{\nu}$ (blue) are
shown. The solid lines show the behavior of the metric potentials in the interior spacetime where as the dashed lines show the nature of the metric coefficients in Schwarzschild geometry. \label{metric}}
\end{figure}
\item {\bf Pressure and density} :~ Fig.~\ref{pr5} depicts the density and pressure profiles. The figures indicate that they are all monotonic decreasing functions of radius `r', having the maximum value at the centre of the star and the pressure $p$ is disappearing at the boundary of the star $r=\mathcal{R}$. At the boundary, however, density is positive.
The central density owing to normal baryonic matter is calculated as follows:
\begin{eqnarray}
\rho_c=\rho(r=0)=\frac{3 a}{8 \pi + 8 \alpha \pi}>0,
 \end{eqnarray}
the central pressure for our present model is obtained as,
 \begin{equation}\label{g1}
 p_c=\frac{a (-1 + 2 \alpha) + 2 (1 + \alpha) B}{8 (1 + \alpha) \pi}>0.
 \end{equation}

We can also see that inside the interior of a star, both pressure and density are non-negative. \\

\begin{figure}[htbp]
        \includegraphics[scale=.55]{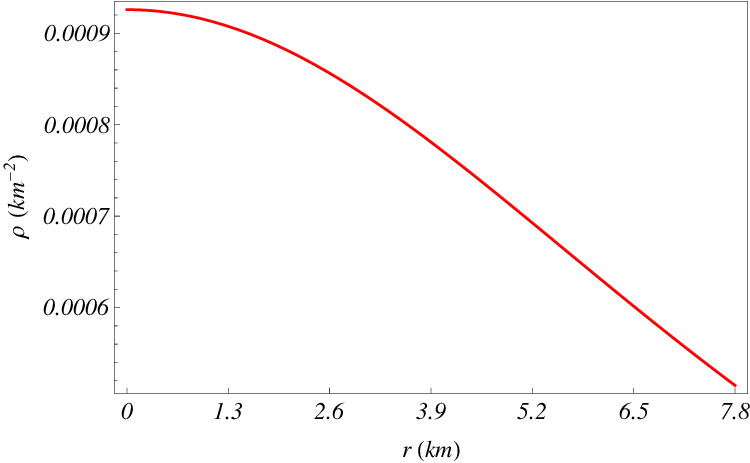}
         \includegraphics[scale=.56]{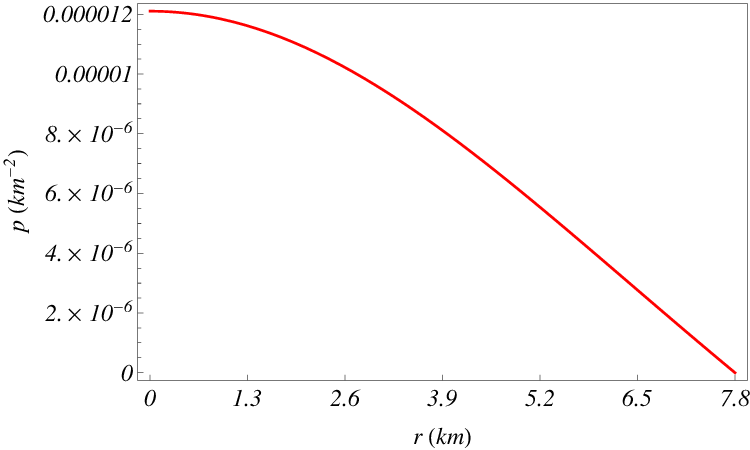}
       \caption{Matter density $\rho$ and pressure $p$ are plotted inside the stellar interior \label{pr5}}
\end{figure}

\item {\bf Zeldovich condition and equation of state parameter} :~The Zeldovich's \cite{Zeldovich} requirement suggests that the pressure to density ratio is smaller than one, i.e., $p/\rho<1$ everywhere within the stellar interior. Now, if we apply the aforementioned condition to the centre of the star, we get  $p_c/\rho_c<1$, which yields the following relationship:
 \begin{eqnarray}\label{g2}
 \frac{a (-1 + 2 \alpha) + 2 (1 + \alpha) B}{3 a} <1,
 \end{eqnarray}
From equations (\ref{g1}) and (\ref{g2}) we get the following relationship,
 \begin{eqnarray}
\frac{1-2\alpha}{2+2\alpha}<\frac{B}{a}<\frac{2-\alpha}{1+\alpha}.
 \end{eqnarray}
 The idea of EoS parameters is explained by the pressure-density ratio, which is symbolised by $\omega$ and calculated as,
 \[\omega=\frac{p}{\rho}=\frac{2 (1 + \alpha) (a + B + (2 b + a B) r^2 + b B r^4)}{
3 a + (a^2 + 5 b) r^2 + 2 a b r^4 + b^2 r^6}-1,\]
It is worth noting that $\omega$ is a crucial factor in determining the type of stellar formations. Now we shall look at the graphical representation of the state parameter $\omega$. It is obvious from Fig. \ref{eos1}, that it is still positive inside the stellar interior, and that the reported value is smaller than $1$. $\omega$ also took zero value at the boundary.
\begin{figure}[htbp]
        \includegraphics[scale=.54]{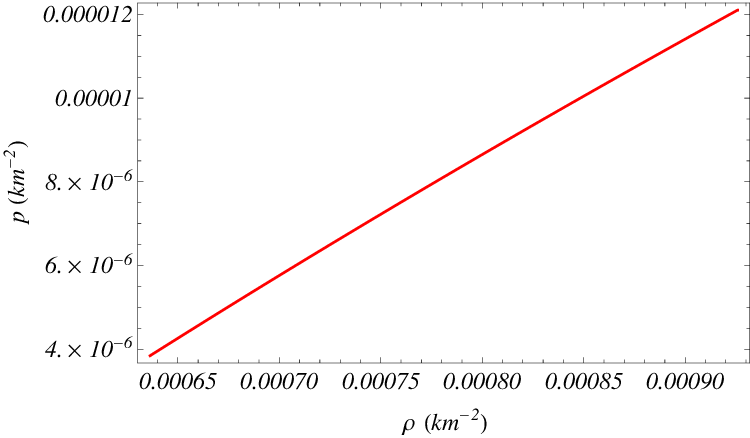}
           \includegraphics[scale=.48]{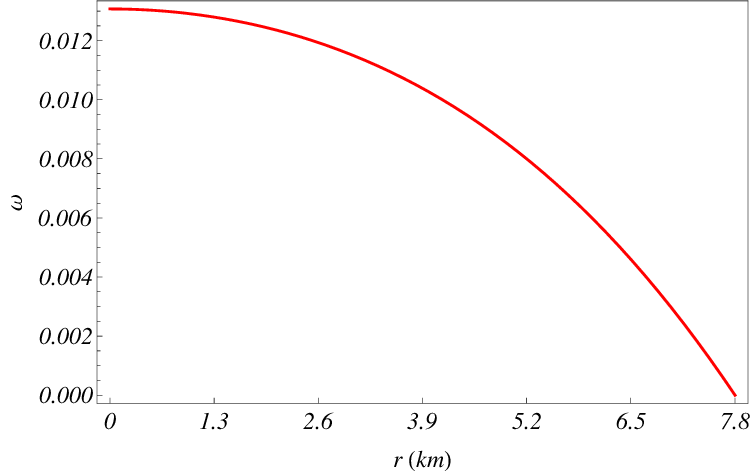}
       \caption{The variation of pressure $p$ with respect to density $\rho$ and the ratio $p/\rho$ are shown inside the stellar interior.\label{eos1}}
\end{figure}

\item {\bf Pressure and density due to dark energy :}~On the other hand, the profiles of pressure and density due to dark energy are depicted in Fig.~\ref{dark1}. The dark energy density $\rho^{de}$ and radial pressure owing to dark energy $p_r^{de}$ are monotonically decreasing functions of `r', whereas transverse pressure corresponding to dark energy $p_t^{de}$ is monotonically growing. Both density and transverse pressure owing to dark energy are positive in our current model, while the radial pressure in agreement with dark energy is negative.

\begin{figure}[htbp]
        \includegraphics[scale=.58]{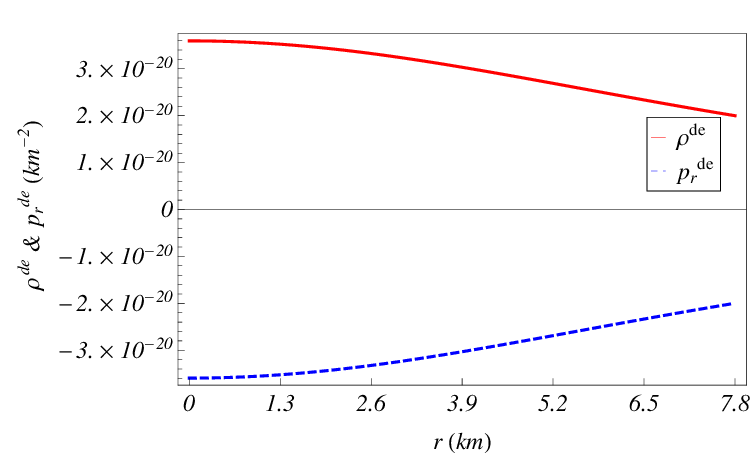}
        \includegraphics[scale=.55]{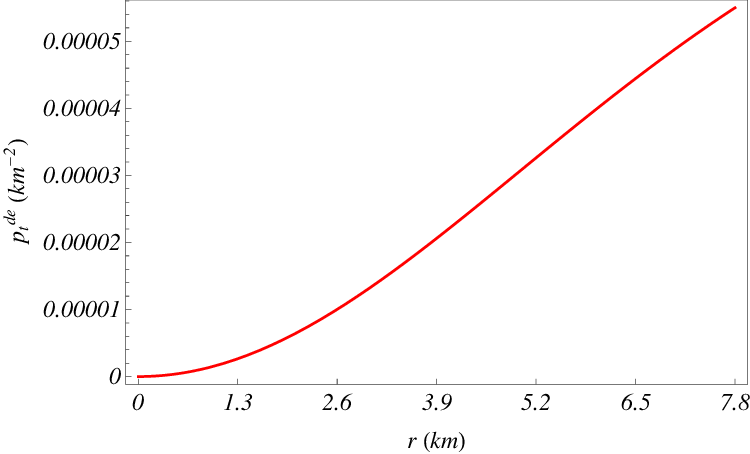}
       \caption{The variation of dark energy density and dark pressure are shown inside the stellar interior.\label{dark1}}
\end{figure}
\end{itemize}
\subsection{Maximality criteria}
 The density and pressure gradient due to the normal baryonic matter for our present model is obtained as,
\begin{eqnarray}
  \frac{d\rho}{dr} &=& -\frac{r (5 a^2 - 5 b + a (a^2 + 13 b) r^2 + 3 b (a^2 + 4 b) r^4 +
    3 a b^2 r^6 + b^3 r^8)}{
 4 (1 + \alpha) \pi (1 + a r^2 + b r^4)^3},\\
  \frac{dp}{dr} &=& \frac{r}{4 (1 + \alpha) \pi (1 + a r^2 + b r^4)^3}\Big[a^2 - 4 a^2 \alpha - b + 4 \alpha b - 2 a B -
 2 a \alpha B + \big(a (a^2 + b - 12 \alpha b)\nonumber\\&&-
    2 (1 + \alpha) (a^2 + 2 b) B\big) r^2 +
 3 b \big(a^2 - 4 \alpha b - 2 a (1 + \alpha) B\big) r^4 +
 b^2 \big(3 a - 4 (1 + \alpha) B\big) r^6 + b^3 r^8\Big] ,
\end{eqnarray}
and \[\rho''(0)= -\frac{5 (a^2 - b)}{4 (1 + \alpha) \pi},\,p''(0)=-\frac{(-1 + 4 \alpha) (a^2 - b) + 2 a (1 + \alpha) B}{4 (1 + \alpha) \pi}.\]\\
\begin{figure}[htbp]
        \includegraphics[scale=.55]{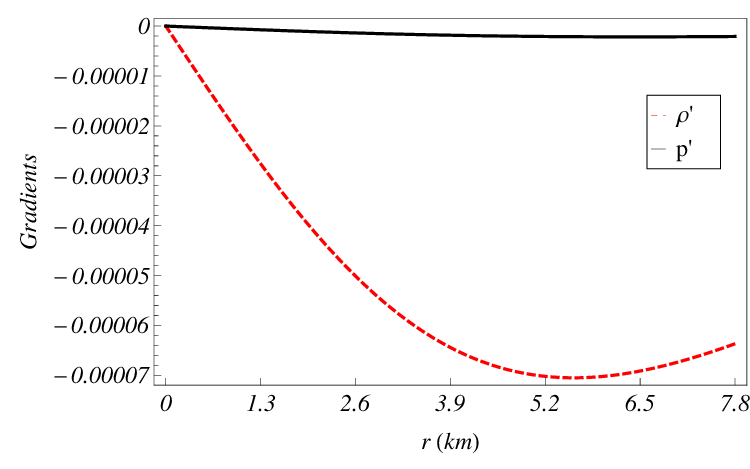}
       \caption{$\frac{d\rho}{dr}$, $\frac{dp}{dr}$ are plotted against $r$ inside the stellar interior.\label{deri}}
\end{figure}
In Fig.~\ref{deri}, we looked at the graphical profiles of both the gradient functions of matter density and pressure. Furthermore, as shown in Fig.~\ref{deri}, the density and pressure gradients stay negative throughout the fluid sphere and vanish at the centre, and second derivatives took negative value at the centre implying that both the density and pressure took maximum value at the centre of the stellar configuration. Table \ref{table2} also represents numerical values for the central density, surface density and central pressure for some well known compact star candidates.

\subsection{Energy conditions}
In this subsection, we examine whether our chosen compact star satisfies the following restrictions for energy conditions inside the boundary. The null energy condition (NEC), weak energy condition (WEC), strong energy condition (SEC) and dominant energy condition (DEC) are the four primary types of energy conditions which are given as follows:
\begin{eqnarray*}
 NEC:~\rho+p\geq 0,\, WEC:~\rho+p\geq 0,~ \rho \geq 0,\, SEC:~\rho+p \geq 0, \rho+ 3p \geq 0,\,DEC:~\rho-p\geq 0,~ \rho \geq 0.
\end{eqnarray*}
The following expressions are required to check the energy conditions :
\begin{eqnarray*}
\rho+p&=&\frac{a + B + (2 b + a B) r^2 + b B r^4}{4 \pi (1 + a r^2 + b r^4)^2},\\
\rho-p&=&\frac{a^2 r^2 -
 a \big(-2 + \alpha + (1 + \alpha) B r^2 - 2 b r^4\big) - (1 + \alpha) B (1 +
    b r^4) +
 b r^2 (3 - 2 \alpha + b r^4)}{4 (1 + \alpha) \pi (1 + a r^2 +
   b r^4)^2},\\
   \rho+3p&=&\frac{3 (B + \alpha (a + B)) + (-a^2 + b + 6 \alpha b +
    3 a (1 + \alpha) B) r^2 + b (-2 a + 3 (1 + \alpha) B) r^4 -
 b^2 r^6}{4 (1 + \alpha) \pi (1 + a r^2 + b r^4)^2}.
\end{eqnarray*}

\begin{figure}[htbp]
        \includegraphics[scale=.55]{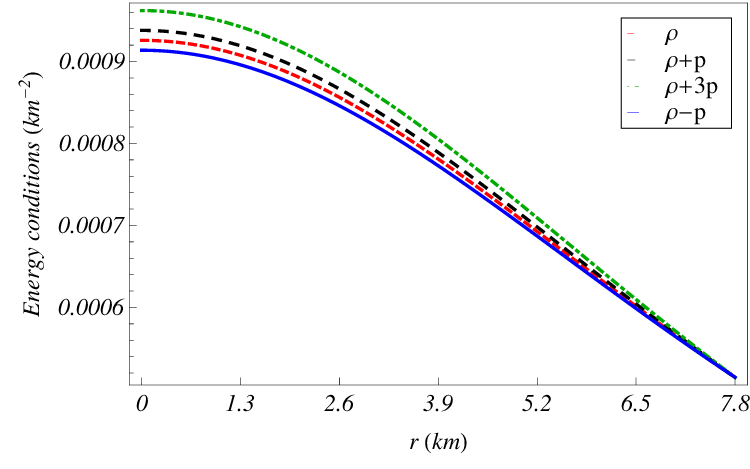}
       \caption{All the energy conditions are plotted inside the stellar interior.\label{ec1}}
\end{figure}
With the help of a graphical representation in Fig.~\ref{ec1}, we have demonstrated that our current model satisfies all of the energy conditions since all the inequalities mentioned above are satisfied by our model.
\subsection{Mass radius relation and surface redshift}
The compactness of the model is determined by a dimensionless parameter $\mathcal{U}$ which is the ratio of maximum mass to the radius and it cannot be arbitrarily large. According to Buchdahl \cite{Buchdahl1959}, the compactness of a model should be smaller than $4/9$ to be a stable structure.
So, to find the compactness factor, in this subsection, we are interested to study the mass function of our model which is the solution of the following differential equation with $m(0)=0$:
\begin{eqnarray*}
\frac{dm(r)}{dr}&=&4\pi\rho(r)r^2 ,
\end{eqnarray*}
Solving the above equation with the given initial condition, we obtain the mass function of the model as,
\begin{eqnarray}
m(r)&=&\frac{r^3(a+br^2)}{2(1+\alpha)(1+ar^2+br^4)}.
\end{eqnarray}
\begin{figure}[htbp]
        \includegraphics[scale=.55]{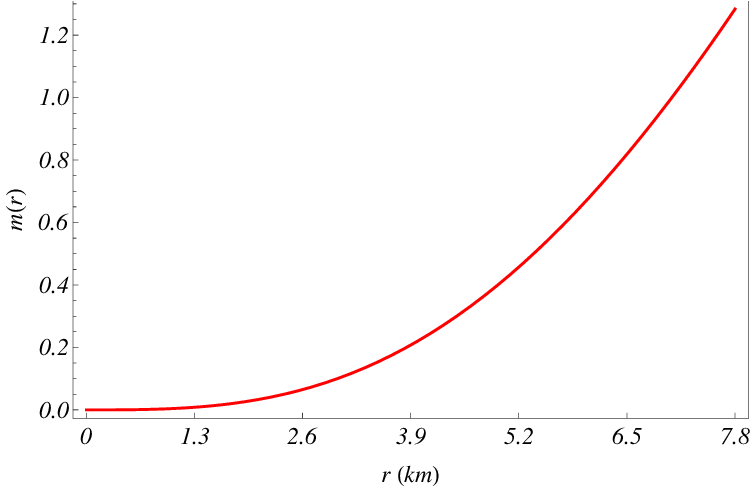}
        \includegraphics[scale=.55]{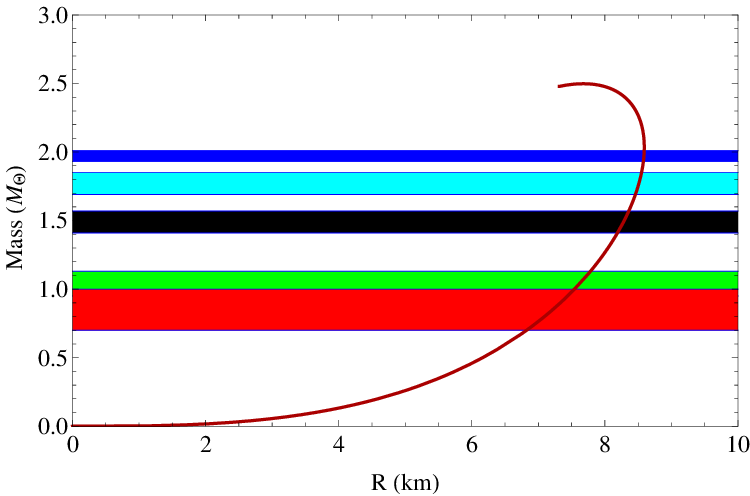}
       \caption{The mass function and $M-R$ relation are plotted inside the stellar interior. The description of the strips is as follows :~(a) Red-Her X-1, (b) Green- LMC X-4, (c) Black- Cen X-3, (d) Cyan- Vela X-1 and (e) Blue- PSR J1614-2230 \label{mr1}}
\end{figure}
It is worth noting that the mass function is influenced by $\alpha$.\\ The compactness factor is calculated as, \[\mathcal{U}=\frac{m(\mathcal{R})}{\mathcal{R}}=\frac{\mathcal{R}^2(a+b\mathcal{R}^2)}{2(1+\alpha)(1+a\mathcal{R}^2+b\mathcal{R}^4)}.\]
Fig.~\ref{mr1} depicts the mass function. The figure shows that the mass function is a monotonic increasing function of `r', meaning that it reaches its highest value at the boundary of the star.
The surface redshift of our present model is calculated as, \[z_s=\frac{1-\sqrt{1-2\mathcal{U}}}{\sqrt{1-2\mathcal{U}}}=\left[1-\frac{\mathcal{R}^2(a+b\mathcal{R}^2)}{(1+\alpha)(1+a\mathcal{R}^2+b\mathcal{R}^4)}\right]^{-\frac{1}{2}}-1.\]
Table~\ref{table2} shows the numerical values for compactness factor and surface redshift for several compact stars. The maximum allowable mass for our present model is shown in Fig.~\ref{mr1}.
\section{Equilibrium and model stability through different tests}\label{equ}
A heuristic treatment is the study of stability mechanisms of the compact structures. There are numerous criteria in the literature that can be used to resolve this problem. In this section, we shall look at several tests that can be used to ensure that our current model is stable and balanced.
\subsection{Velocity of sound}
In this subsection we are interested in calculating the sound velocity of our present model which reflects the stiffness of the system. According to the definition, the sound velocity of our system is calculated as follows:
\begin{eqnarray*}
V_s=\sqrt{\frac{dp}{d\rho}}=\sqrt{\frac{p'}{\rho'}},
\end{eqnarray*}
From our current model, we can calculate the square of sound velocity as,
\begin{eqnarray}
V_s^2&=&\frac{dp}{d\rho} =\frac{1}{5 a^2 - 5 b + a (a^2 + 13 b) r^2 + 3 b (a^2 + 4 b) r^4 + 3 a b^2 r^6 +
  b^3 r^8}\times \Big[-a^2 + 4 a^2 \alpha + b - 4 \alpha b \nonumber\\&&+ 2 a B +
 2 a \alpha B + (-a (a^2 + b - 12 \alpha b) +
    2 (1 + \alpha) (a^2 + 2 b) B) r^2 +
 3 b (-a^2 + 4 \alpha b + 2 a (1 + \alpha) B) r^4 \nonumber\\&&+
 b^2 (-3 a + 4 (1 + \alpha) B) r^6 - b^3 r^8\Big].
\end{eqnarray}
\begin{figure}[htbp]
        \includegraphics[scale=.55]{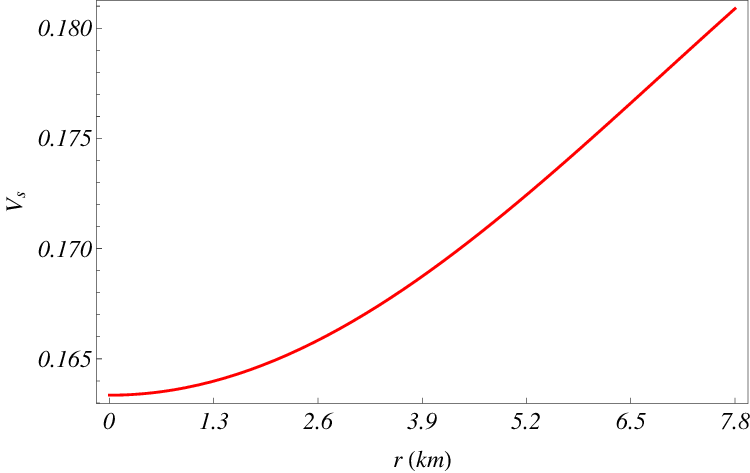}
       \caption{ The sound velocity is shown against `r' inside the stellar interior \label{sv1}}
\end{figure}

Fig.~\ref{sv1} depicts the $V_s$ profile. It is familiar that the slope of the $p(\rho)$ function determines the sound velocity. In principle, sound should travel at a slower rate than light, and a lower sound velocity correlates to a softer equation of State (EoS). We can conclude from the graphical analysis that $V_s$ is less than $1$. As a result, the causality criterion is met for our model.

\subsection{TOV equation}
In this subsection, we shall look at the equilibrium for this present model in presence of dark energy. To check the equilibrium of the model under various forces acting on it, we shall use the Tolman-Oppenheimer-Volkoff (TOV) equation, which is described by \cite{Ghezzi:2009ct},
 \begin{eqnarray}\label{g7}
\frac{\nu'}{2}(\rho+p)+\frac{2}{r}(p_r^{\text{de}}-p_t^{\text{de}})+\frac{d}{dr}(p_r^{\text{eff}})=0,
\end{eqnarray}
where $p_r^{\text{eff}}$ is the total radial pressure. Rearranging the terms of the equation (\ref{g7}), one can obtain the following equation :
\begin{eqnarray}
\frac{d}{dr}(p_r^{\text{eff}})&=&-\frac{\rho+p}{2}\frac{\left(m+4\pi p_r^{\text{eff}}r^3\right)}{r(r-2m)}+2\frac{\Delta P}{r},
\end{eqnarray}
where $\Delta P=p_t^{\text{de}}-p_r^{\text{de}}$ and $m(r)=4\pi\int_0^{r}(\rho+\rho^{de})r^2 dr$. For a constant dark energy density: $m(r) =M(r)+\frac{1}{6}\Lambda r^3$, where $M(r)=4\pi\int_0^{r}\rho r^2 dr$. Bowers and Liang \cite{Bowers1974} first studied the above equation and its solution. In terms of dark energy density the above equation can be written as :
\begin{eqnarray}
\frac{d}{dr}(p_r^{\text{eff}})&=&-\frac{(\rho+p)}{2}\frac{\left\{M(r)+4\pi p r^3-\frac{8\pi}{3}\rho^{\text{de}}r^3\right\}}{r\left\{r-2M(r)-\frac{8\pi}{3}\rho^{\text{de}}r^3\right\}}+2\frac{\Delta P}{r}.
\end{eqnarray}
The above equation can be written as a function of the cosmological constant $\Lambda=8\pi \rho^{\text{de}}$ as,
\begin{eqnarray}
\frac{d}{dr}(p_r^{\text{eff}})&=&-\frac{(\rho+p)}{2}\frac{\left\{M(r)+4\pi p r^3-\frac{1}{3}\Lambda r^3\right\}}{r\left\{r-2M(r)-\frac{1}{3}\Lambda r^3\right\}}+2\frac{\Delta P}{r}.
\end{eqnarray}
For $\Lambda=0$ and $\Delta P=0$, this equation reduces to the well known Tolman-Oppenheimer-Volkov (TOV) equations \cite{Oppenheimer:1939ne}.\\
The equation (\ref{g7}) can be written as,
\begin{eqnarray}
F_g+F_h+F_d=0,
\end{eqnarray}
The expressions of three different forces mentioned in the above equation are given as follows :
\begin{eqnarray*}
\text{Gravitational force :~~} F_g=  -\frac{1}{2}\nu'(\rho+p),\\
\text{Hydrostatic force :~~} F_h=-\frac{dp}{dr},\\
\text{ Force due to dark energy :~~} F_d=\frac{2}{r}(p_t^{\text{de}}-p_r^{\text{de}})-\frac{d}{dr}p_r^{de}.
\end{eqnarray*}

\begin{figure}[htbp]
        \includegraphics[scale=.55]{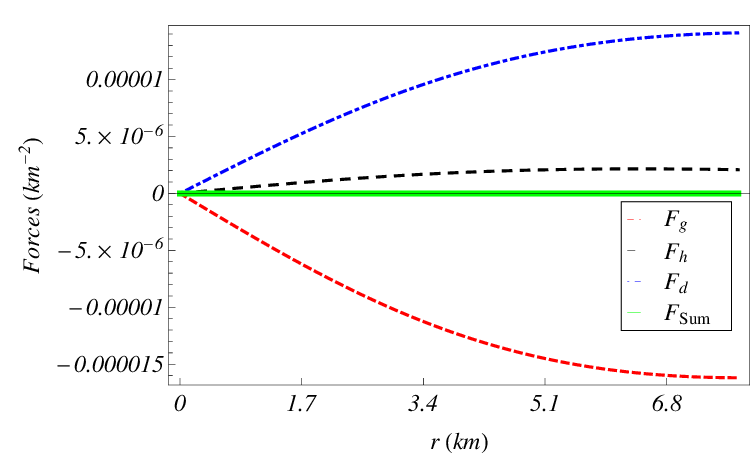}
       \caption{Different forces acting on the present model are plotted against $r$ for the Hybrid star Her X-1 by taking different values of $\alpha$.\label{force1}}
\end{figure}
Fig.~\ref{force1} illustrates the graphical representation of all these forces. It should also be noted that the hydrostatic and dark energy forces are positive, whereas the gravitational force is negative. From the plots of all of these forces, it is clear that the overall effect of these forces $F_{Sum}=F_g+F_h+F_d$ vanishes, as seen in the green plot. The balance of our model is therefore achieved.
\subsection{Harrison-Zeldovich-Novikov condition}

\begin{figure}[htbp]
        \includegraphics[scale=.55]{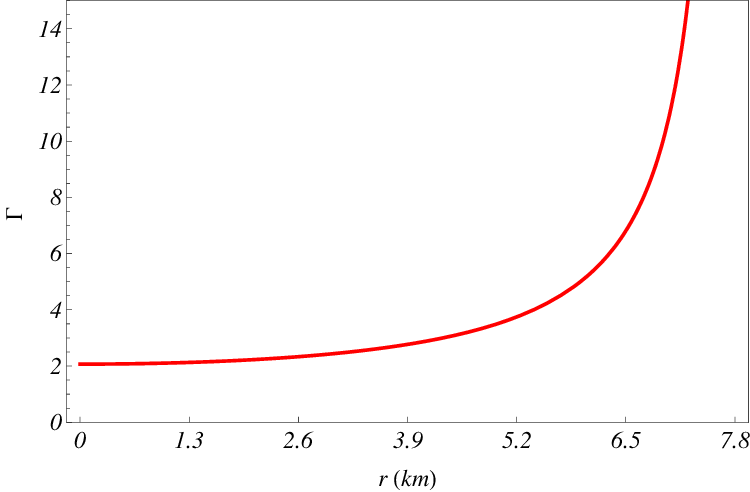}
         \includegraphics[scale=.62]{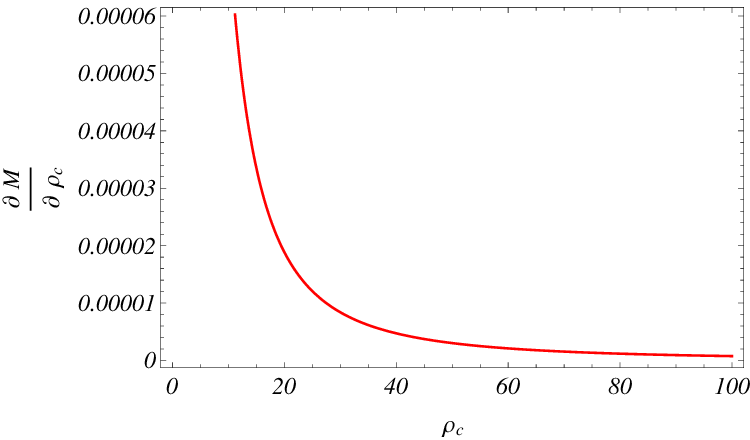}
       \caption{The relativistic adiabatic index and $\frac{\partial M}{\partial \rho_c}$ are shown inside the stellar interior \label{stab1}}
\end{figure}

The Harrison-Zeldovich-Novikov \cite{Zeldovich,Harrison1965} criterion for the stability of the self-gravitating compact star tells that the mass of the star increases with central density in the stable region, which mathematically entails $\frac{\partial M}{\partial \rho_c}>0$. For our present model,
 \begin{eqnarray}
 M=\frac{\mathcal{R}^3}{2(1+\alpha)}\frac{8\pi \rho_c (1+\alpha)+3b\mathcal{R}^2}{3+8\pi \rho_c (1+\alpha)\mathcal{R}^2+b\mathcal{R}^4}.
 \end{eqnarray}
 Fig.~\ref{stab1} shows that, though $\frac{\partial M}{\partial \rho_c}$ becomes monotonically decreasing as $\rho_c$ increases, it is always positive throughout the stellar structure. So, according to the Harrison-Zeldovich-Novikov condition, our model is stable and physically realistic.

\subsection{Relativistic adiabatic index}

The stiffness of the equation of state for a given density is represented by the ratio of the two specific heats, known as the adiabatic index $\Gamma$. This adiabatic index $\Gamma$ can be used to investigate the dynamical stability of the stellar structure against an infinitesimal radial adiabatic perturbation. According to the literature review \cite{bondi1964contraction,Chan1993,Chandrasekhar:1964zz}, the adiabatic index for a stable Newtonian sphere should be greater than $4/3$ inside a dynamically stable fluid distribution. For our system, the expression of adiabatic index is given by,
\begin{eqnarray*}
\Gamma&=&\left(1+\frac{\rho}{p}\right)V_s^2,\\
&=&\frac{2 (1 + \alpha) \left(a + B + (2 b + a B) r^2 + b B r^4\right)}{2 a \alpha -a+
 2 B + 2 \alpha B - \left(a^2 + b - 4 \alpha b - 2 a (1 + \alpha) B\right) r^2 -
 2 b \left(a - (1 + \alpha) B\right) r^4 - b^2 r^6}V_s^2.
\end{eqnarray*}
Due to the complexity of the expression, we used graphical analysis to assess the stability of our current model under the adiabatic index $\Gamma$. $\Gamma$ is greater than $4/3$ everywhere throughout the stellar interior, as shown in Fig~\ref{stab1}.

\section{Discussion}

Motivated by recent evidences regarding the accelerated expansion of our universe, during the last few decades, many researchers have focused in developing dark energy models which is a suitable candidate to explain this phenomenon. In reality, dark energy has opened up new avenues for theoretical cosmology and astrophysics. The stellar model in this work is developed by using the TK {\em ansatz} and is based on a mixing of two fluids. We choose 4U $1538-52$ as a typical of compact star candidates for numerical analysis and physical justifications of the obtained solutions. It has an observed mass of $(0.87 \pm 0.07)M_{\odot}$ and radius $7.866 \pm 0.21$ km. The numerical values of the constants present in the metric coefficients were determined from the matching condition as well as the numerical values of other thermodynamical parameters are also found for a total of $10$ compact stars. Tables \ref{table1} and \ref{table2} contain all of the numerical results. We have presented a comparison of the obtained results with the observational constraints, and found that our model matches the observational values.\par
Fig.~\ref{metric} depicts the nature of the metric potential inside the star interior, note that both of them are singularity-free and regular. Since compact relativistic objects are the densest, the impact of matter density and pressure at the core of the star should be maximum. Fig.~\ref{pr5} depicts the variation in matter density and pressure owing to normal matter in the interior of a compact star with respect to radial coordinate `r'. These graphs illustrate that at the centre of the dark energy star, both density and pressure are at their highest levels, indicating the presence of highly compact cores. It is also found that the pressure vanishes at the boundary $r=\mathcal{R}$, while the density and pressure components are monotonically decreasing functions of radial coordinate `r'. The relation between pressure and density is depicted in Fig.~\ref{eos1}, and it can be seen that they follow a linear equation of state. We examined the equation of state parameter $\omega=p/\rho$, which remained between $0$ and $1$, indicating that the matter content is non-exotic in nature, which is still another indicator of the favorable behavior of the obtained solution. The behavior of the density and pressure components owing to dark energy, on the other hand, is depicted in Fig.~\ref{dark1}, and it is evident that $p_r^{de}$ is negative while $\rho^{de}$ is positive, which is a common feature of dark energy stars. The negative behavior of pressure and density gradients are shown in Fig.~\ref{deri}. Fig.~\ref{ec1} shows that our system is consistent under all energy situations, proving the physical validity of the obtained solutions. In the absence of charge, the highest permissible mass-to-radius ratio for a compact star model should not exceed the Buchdahl limit \cite{Buchdahl1959}, i.e., $\mathcal{U}<4/9$. On the other hand, the value of surface redshift, cannot be arbitrarily large. In the absence of a cosmological constant $\Lambda$, it is found that the surface redshift $z_s$ is in the range $z_s \leq 2$ \cite{Buchdahl1959,bohmer2006bounds,straumann2012general}. We estimated the compactness factor and surface redshift of many compact stars using our current model and presented their values in Table \ref{table2}. All of the numerical values in the table are within the predicted range. Using the Tolman-Oppenheimer-Volkoff equation, we were able to achieve hydrostatic equilibrium in our system. It can be observed that the system is governed by three forces: gravitational ($F_g$), hydrostatic $(F_h)$, and dark energy forces ($F_d$). We discovered that a single attracting force $F_g$ counterbalanced the combined effect of hydrostatics and dark energy force. As a result, the net effect of applied forces on the system disappears, and the system is in equilibrium. The relativistic adiabatic index and Harrison-Zeldovich-Novikov criteria are also used to describe the stability of the model.\par
In this respect we want to mention some previous works related to dark energy star. Valera et al. \cite{Varela:2010mf} previously discovered a link between their construction and a charged strange quark star, as well as models of dark matter that include massive charged particles. Rahaman et al. \cite{Rahaman:2010mr} suggested the existence of a Chaplygin charged dark energy star or a strange quark star with a radius of roughly $8$ km using a Chaplygin-type EOS. Rahaman et al. \cite{Rahaman:2011hd} obtained a singularity-free spherically symmetric dark energy star in the KB platform with a radius of $15$ km such that both $p_t^{\text{eff}}$ and $p_t^{\text{shell}}$ concurrently vanishes at $r=15$ km. The model physically incorporates anisotropic matter restricted within $8$ km of the center of the star, as well as a thick shell extending up to $7$ km and characterized by zero energy density and non-zero transverse pressure in the outer region. Hopefully, in the future, this type of theoretical modeling will receive observational support. {\bf In $2016$, Bordbar et al. \cite{Bordbar:2015wva} proposed neutron stars model by considering d-dimensional (d $\geq$ 3) spherically symmetric line element in the context of Einstein-$\Lambda$ gravity inspired by the observations indicating that the universe is expanding at an accelerating rate. The cosmological constant is the most basic form of dark energy, leading to the current standard model of cosmology, known as the $\Lambda$-CDM model, which fits many cosmological observations successfully. In this work, the authors used contemporary equations of state of neutron-star matter derived from microscopic simulations to investigate the maximum mass of neutron stars. The results reveal that the cosmological constant has an impact on the maximum mass of neutron star. In other words, when the value of $\Lambda$ is increased, the maximum mass of neutron stars decreases (choosing $\Lambda>0$). The behavior of diagrams of mass versus radius of neutron stars is one of the fascinating conclusions obtained in this paper. These diagrams demonstrated a transition from a neutron star to a quasi-quark star by increasing $\Lambda$. Their findings revealed that when the estimated value of cosmological constant was $10^{-52} m^{-2}$, the constant had no discernible effect on the structure of neutron stars. They also investigated that the maximum mass and radius of the neutron star were reduced by using larger values of  $\Lambda$ (about $\Lambda> 10^{-14} m^{-2}$). The effect of the cosmological constant on the so-called gravity strength is another noteworthy conclusion of this article. It was also examined that when the positive value of the cosmological constant $\Lambda$ increases, the strength of gravity diminishes, and this effect leads to a decreasing of the maximum mass of neutron stars.}\par
In conclusion, we obtained a singularity-free and stable generalized model using the dark energy equation of state, which is ideal for analysing dark energy stars in the Einstein gravity framework.

\section*{Acknowledgements} P.B is thankful to IUCAA, Govt of India, for
providing visiting associateship.

\bibliography{hybrid1}

\end{document}